\documentclass[reprint,aps,prl,twocolumn,superscriptaddress]{revtex4-1}
\usepackage{braket}
\usepackage{amssymb}
\usepackage{amsmath}
\usepackage{epsfig}
\usepackage{color}
\usepackage{graphics, graphicx}
\usepackage{bbold}
\usepackage{psfrag}
\usepackage{mathcomp}
\usepackage{subfigure}
\usepackage{verbatim}
\usepackage[colorlinks, citecolor=blue]{hyperref}
\usepackage[normalem]{ulem}
\usepackage{bm}

\begin{document}

\title{Reply to the Comment on ``Shell-Shaped Quantum Droplet in a Three-Component Ultracold Bose Gas"}
\author{Yinfeng Ma}
\affiliation{Beijing National Laboratory for Condensed Matter Physics, Institute of Physics, Chinese Academy of Sciences, Beijing 100190, China}
\affiliation{Department of Basic Courses, Naval University of Engineering, Wuhan 430033, China}
\author{Xiaoling Cui}
\affiliation{Beijing National Laboratory for Condensed Matter Physics, Institute of Physics, Chinese Academy of Sciences, Beijing 100190, China}

\maketitle


In our Letter\cite{Ma}, we proposed a self-bound shell-shaped BEC in a three-component ($1,2,3$) Bose gas, where $(2,3)$ and $(1,2)$ droplets are linked as core-shell structure. A recent Comment\cite{Comment} argued that a ``dimer" configuration should be instead the ground state, where $(2,3)$ and $(1,2)$ stay side-by-side. Moreover, \cite{Comment} also explored the state formation, finding that a naive trap-release protocol was unable to produce the core-shell structure. In this reply we show that our core-shell structure is an excited state for finite-size systems, while it becomes energetically degenerate with dimer configuration in thermodynamic limit. Furthermore, we find the core-shell structure is locally stable under external perturbations, and if one pays careful attention to mode-matching, a trap-release protocol can well produce this structure.

\begin{figure}[b]
    \centering
    \includegraphics[width=9cm]{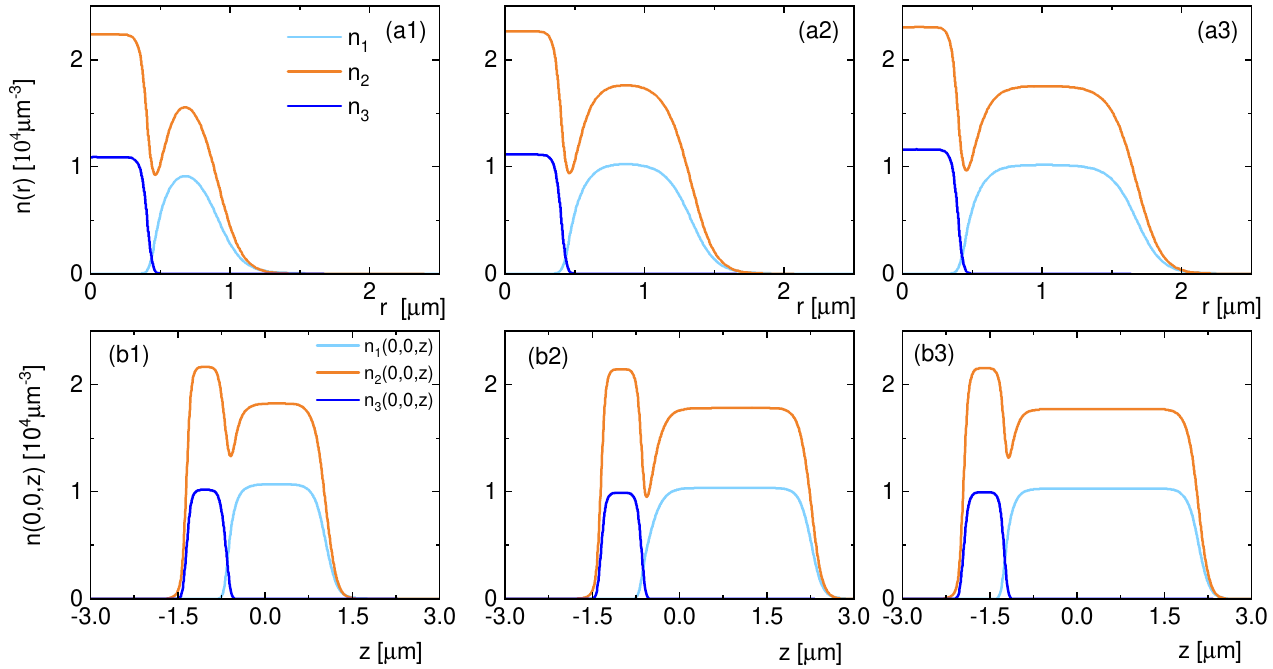}
    \caption{Density profiles of core-shell (a1,a2,a3) and dimer (b1,b2,b3) states. The atoms numbers are $(N_1,N_2,N_3)/10^5=(0.3,0.56,0.03)$ in (a1,b1), $(1,1.76,0.03)$ in (a2,b2) and $(2,3.47,0.03)$ in (a3,b3). (a1,a2,a3) are for radial densities, and (b1,b2,b3) are densities at $x=y=0$. The relative energy difference, defined as $\delta E\equiv (E_{\rm c-s}-E_{\rm dimer})/|E_{\rm c-s}|$, decreases as the shell atom number grows: $\delta E=13.36\%$(a1,b1), $6.0\%$(a2,b2) and $3.12\%$(a3,b3). Here we consider a realistic $^{23}$Na-$^{39}$K-$^{41}$K ('1'-'2'-'3') mixture near $B\sim 150$G with $a_{23}=-200a_0$.  
}\label{fig_compare}
\end{figure}

The equilibrium core-shell and dimer configurations can be obtained by choosing different initial states in the energy minimization process, namely, the former (latter) is from a rotationally invariant (symmetry broken) initial state\cite{footnote_symmetry}. Fig.\ref{fig_compare} shows their typical density profiles with varying atom numbers. We find that the core-shell state indeed has a higher energy than the dimer one for finite-size systems, with energy cost mainly from  the surface of the shell. As the shell becomes larger, however, their relative energy difference decreases (see Fig.\ref{fig_compare}),  due to less contribution from the surface  as compared to the bulk.  In the thermodynamic limit with vanishing surface contribution, the two states become degenerate in energy per particle, which is solely determined by equilibrium densities of correlated $(1,2)$-$(2,3)$ droplets as derived in our Letter\cite{Ma}.

\begin{figure}[t]
    \centering
    \includegraphics[width=7cm, height=5cm]{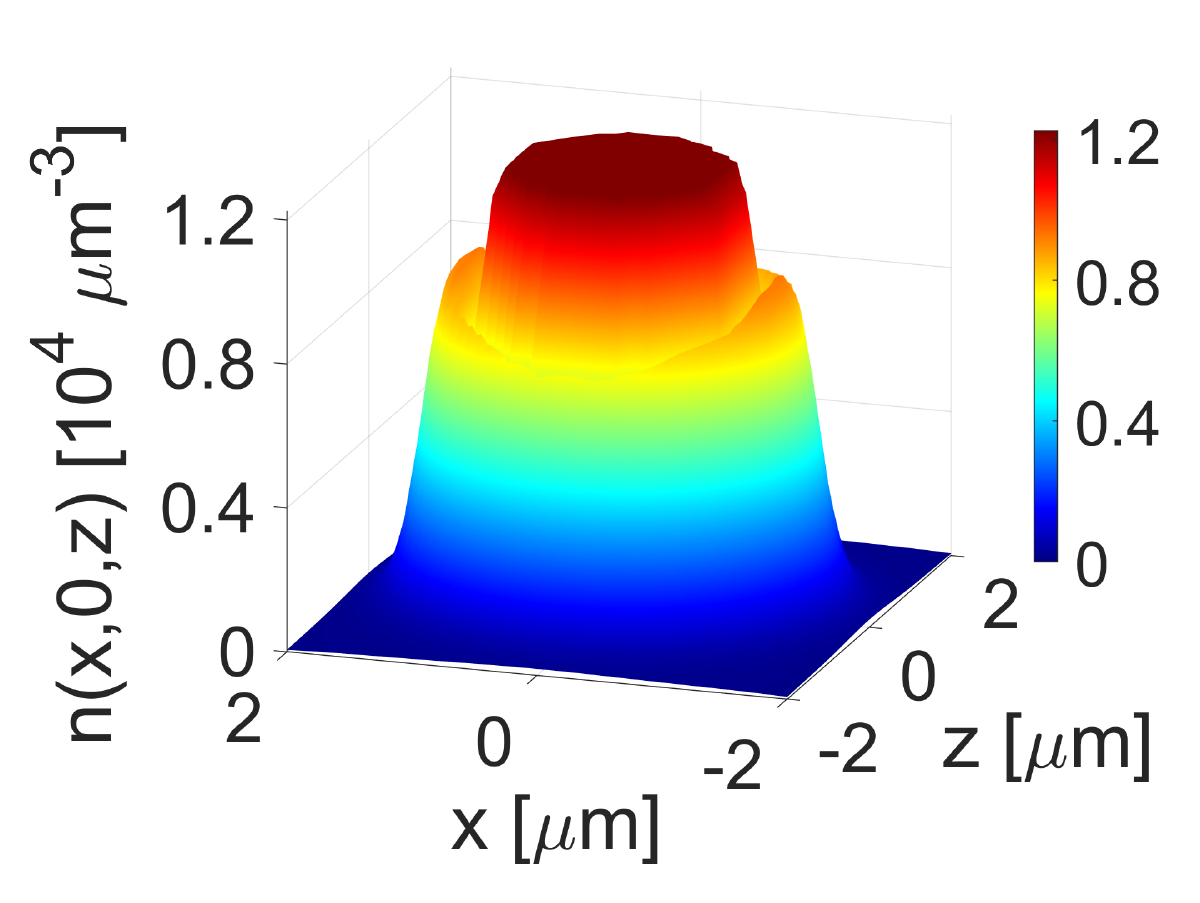}
    \caption{Deformation of core-shell structure under a magnetic field gradient $B'=0.08 E_0/l_0$ ($l_0=1\mu m$ and $E_0=\hbar^2/(m_Kl_0^2)$). Here we take the initial core-shell state as in Fig.2(a3) of \cite{Ma}. To clearly see the deformation we just plot out $n_1$ (shell component) and $n_3$ (core component) at $y=0$; $n_2$ (not shown here)  exhibits similar core-shell structure. }\label{fig_B}
\end{figure}

The stability of core-shell state can be inferred from its all-real excitation spectra, as reported in Fig.4 of \cite{Ma}. Should the system be unstable against any density fluctuation, its associated mode will develop imaginary part and grow exponentially with time to destabilize the system. For all the atom numbers considered in our work, we do not observe such instability.  To further confirm the stability of core-shell structure, we add to the system a small symmetry-breaking field as the magnetic field gradient, $h_{sb}(\{{\bf r_i}\})=B'(z_3-z_1)$ with $B'(>0)$\cite{footnote_B}. Note that $h_{sb}$ particularly favors the dimer state where $(1,2)$ stays at $z_1>0$ and $(2,3)$ at $z_3<0$. We consider the thin shell case in Fig.2(a3) of our work\cite{Ma}, which was commented in \cite{Comment}  as very unstable and easily decaying to dimer state with a small amount of perturbation. However, we observe that the shell structure is only slightly deformed under a small $B’=0.08 E_0/l_0$, rather than ending up with dimer state (here $l_0=1\mu m$ is the typical length scale of the system and $E_0=\hbar^2/(m_Kl_0^2)$). The failure of transiting to dimer state directly manifests the intrinsic stability of core-shell structure --- under perturbation it can adjust its density profile such that the surface tension can reach a balance with external force and help to resist further deformation of the structure.   The common component-2 also plays a positive role here, whose core-shell structure efficiently glues 1 and 3 as similar shape.  Still, it is possible that the thermal or high-order quantum fluctuations
(beyond those captured by our extended GP ansatz) could destabilize the core-shell structure. Nonetheless our current calculations strongly suggest that it is observable.

\begin{figure}[t]
    \centering
    \includegraphics[width=9cm]{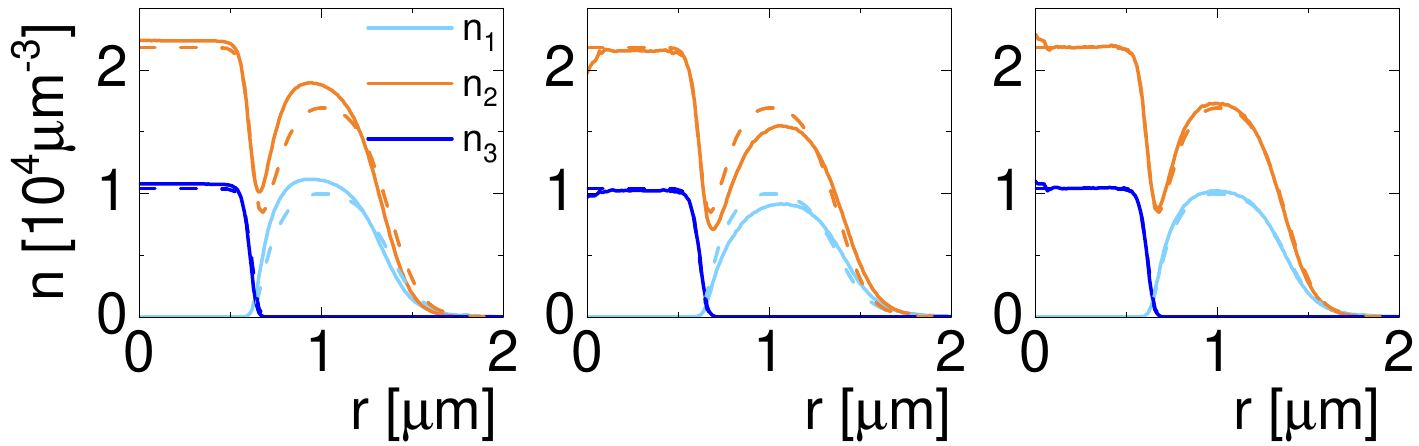}
    \caption{Time evolution of radial densities after the system is released from an isotropic harmonic trap. Here we take the trap frequency as $\omega_{\rm K}=5KHz$, and $\omega_{\rm Na}=\omega_{\rm K}\sqrt{m_{\rm K}/m_{\rm Na}}$; the atom numbers are $(N_1,N_2,N_3)/10^5=(1,1.86,0.1)$, and the time is $t\ (ms)=0$(a), $0.15$(b) and $0.3$(c). Dashed lines with according color show the densities of equilibrium core-shell state in free space. The interaction parameters are the same as in Fig.\ref{fig_compare}.} \label{fig_trap}
\end{figure}

According to the scheme in \cite{Ma}, one first prepares the core-shell structure initially in an isotropic harmonic trap, which does not favor the dimer state, and then release the system from the trap.  The system is expected to quickly relax to the ground state core-shell structure under an optimized mode-matching, as successfully implemented in recent experiment\cite{Guo2021}. We have confirmed the feasibility of this scheme in Fig.\ref{fig_trap}. At time $t=0$, the trapped system has an obviously higher shell densities than the free space case (Fig.\ref{fig_trap}(a)). After releasing from the trap $(t>0)$, the system quickly relaxes, and both core and shell end up with breathing oscillations around  free-space profiles, see Fig.\ref{fig_trap}(b,c).  During the whole process, the system preserves rotational symmetry and we do not observe strange geometry as described in \cite{Comment}. The core-shell structure is also found to be robust under small external perturbations, for the same reason as in equilibrium case. 

We emphasize that the mode-matching is very important for achieving shell-shaped BEC in above process, which requires that the density profiles of initial  (in a trap) and final (without trap) states match each other as much as possible.
If this requirement is violated, there will be a huge amount of internal energy to release during the dynamics, leading to complex dynamical outcomes.
In fact, \cite{Comment} did not follow the mode-matching scheme as ours --- the initial and final states therein have opposite core and shell components\cite{footnote_trap}. As a result, during the releasing dynamics $(1,2)$ and ($2,3$) tend to flow oppositely, and the large initial energy can also result in splash dynamics of the whole system\cite{Ma2}. Together with symmetry-breaking perturbations, splitting small droplets with strange geometry may be resulted\cite{Comment}.


\begin{thebibliography}{99}
\bibitem{footnote_symmetry} The core-shell state always preserves $SO(3)$ rotational  symmetry during the real or imaginary time evolution, given the same symmetry of total Hamiltonian. In comparison, the dimer structure is from a symmetry-broken initial state where $(1,2)$ and $(2,3)$ droplets are arranged side-by-side.
\bibitem{footnote_B}  Note that a field  gradient on component-2 can be gauged away in the moving frame of this component, so only the field gradients on two of the components (here $1$ and $3$) are independent.
\bibitem{footnote_trap} The switched core and shell components may be due to the combined effect of tight trapping potential and large $N_{1,2}$ taken in \cite{Comment}, such that $(1,2)$ prefers to stay in the trap center (core) to avoid large potential energy and then $(2,3)$ becomes shell.
\end{thebibliography}
\end{document}